\documentclass{article}
\usepackage{spconf,amsmath,graphicx,hyperref}
\usepackage{amsfonts}
\usepackage{amsmath}
\usepackage{booktabs}
\usepackage[table]{xcolor}
\usepackage{colortbl}

\title{Temporally Heterogeneous Graph Contrastive Learning for Multimodal Acoustic Event Classification}
\name{Yuanjian Chen$^{\star}$ \qquad Yang Xiao$^{\dagger}$\thanks{$^{\dagger}$Corresponding author (email: yxiao9550@student.unimelb.edu.au)} \qquad Jinjie Huang$^{\star}$}
\address{$^{\star}$Harbin University of Science and Technology \\ $^{\dagger}$The University of Melbourne}
\begin{document}
%
\maketitle
\begin{abstract}
Multimodal acoustic event classification plays a key role in audio-visual systems. Although combining audio and visual signals improves recognition, it is still difficult to align them over time and to reduce the effect of noise across modalities. Existing methods often treat audio and visual streams separately, fusing features later with contrastive or mutual information objectives. Recent advances explore multimodal graph learning, but most fail to distinguish between intra- and inter-modal temporal dependencies. To address this, we propose Temporally Heterogeneous Graph-based Contrastive Learning (THGCL). Our framework constructs a temporal graph for each event, where audio and video segments form nodes and their temporal links form edges. We introduce Gaussian processes for intra-modal smoothness, Hawkes processes for inter-modal decay, and contrastive learning to capture fine-grained relationships. Experiments on AudioSet show that THGCL achieves state-of-the-art performance.
\end{abstract}
\section{Introduction}
\label{sec:intro}

Multimodal acoustic event classification (AEC) ~\cite{atilgan2018integration,xiao2024wilddesed} is an important task in intelligent audio-visual systems. It supports many real-world applications, including security monitoring, multimedia content retrieval, and human–computer interaction~\cite{ghafir2016survey}. These systems benefit from combining audio and visual signals to understand complex environments. However, in practical scenarios, audio signals are often unclear due to background noise, overlapping sounds, or poor recording conditions. Relying on audio alone may lead to errors in event recognition~\cite{ma2021active}. To address this issue, visual cues such as object movement, scene transitions, or lip motion can provide valuable complementary information. As a result, integrating both modalities can improve recognition performance.

While this multimodal approach shows great promise, it also introduces new challenges. One of the key difficulties lies in modeling the correct timing between audio and visual inputs~\cite{dupont2000audio,yin2025exploring,chen2025noise}. Events often follow a strict temporal sequence, and even small misalignments between modalities can confuse the model. Therefore, it is essential to design systems that can effectively capture temporal structures across both audio and visual data. Most existing multimodal methods process audio and visual features separately before combining them. Typically, each modality is encoded by a dedicated neural network, and its features are fused later through concatenation~\cite{liu2021audio,oorloff2024avff,xiao2024mixstyle}. To reduce the noise across modalities, many researchers also introduce additional learning objectives such as contrastive loss or mutual information maximization~\cite{ma2021active,saeed2021contrastive}. For example, cross-modal teacher-student frameworks~\cite{owens2016ambient} learn more robust embeddings by encouraging agreement between audio and visual signals. Other methods such as XDC~\cite{alwassel2020self} and Evolving Losses~\cite{piergiovanni2020evolving} have demonstrated the benefit of combining single-modal and multimodal pretext tasks to improve representation learning.


\begin{figure*}[t!]
\centering
\includegraphics[width=0.7\textwidth]{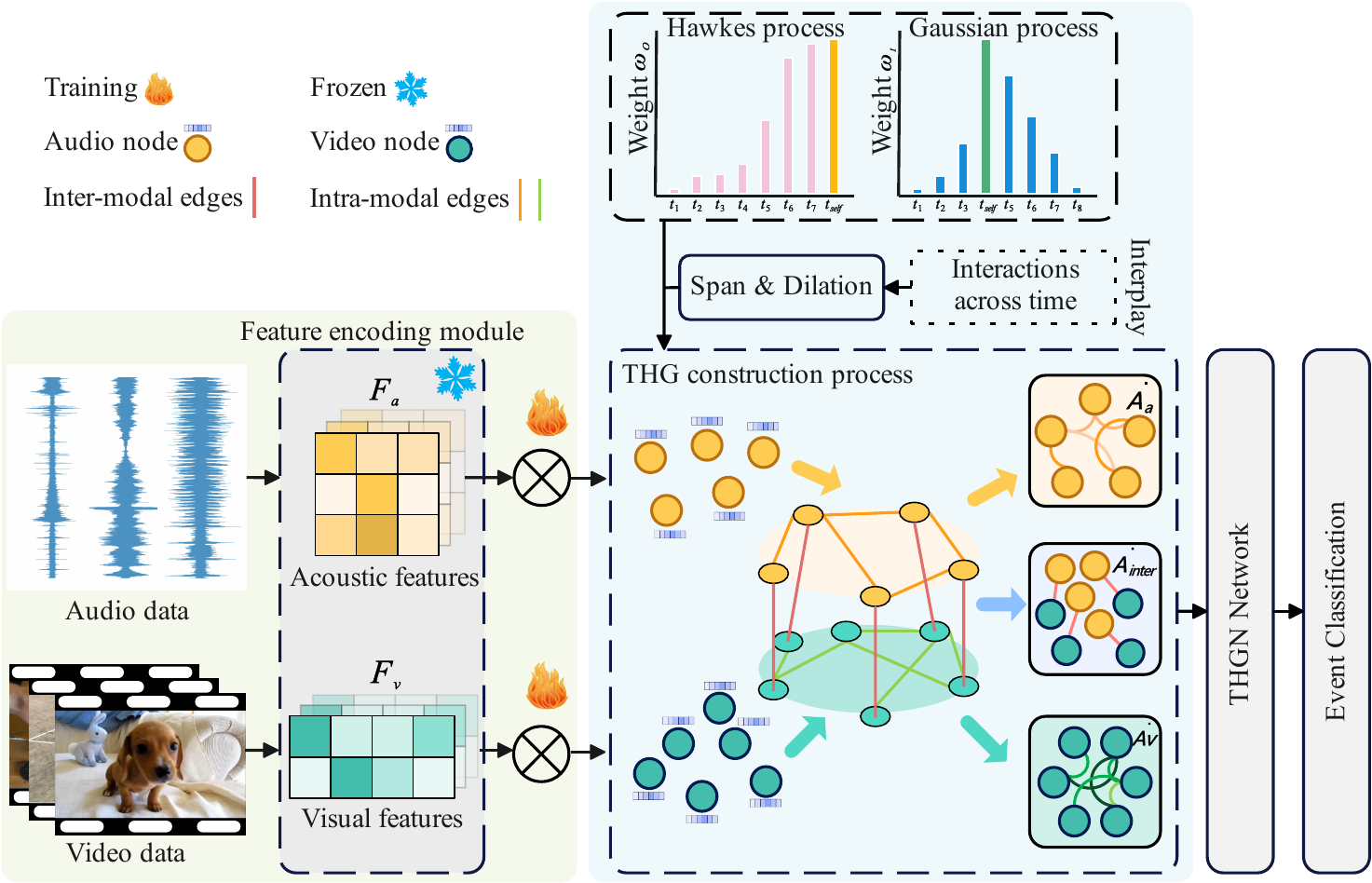}
\vspace{-8pt}
\caption{The overall framework of Temporally Heterogeneous Graph-based Contrastive Learning (THGCL)}
\vspace{-15pt}
\label{fig:fig1}
\end{figure*}

Beyond feature fusion, graph-based learning offers a promising path for modeling multimodal relationships~\cite{yang2023cluster}. Although still uncommon in AEC, such approaches have achieved success in other areas. For example, graphs have been used to link image-text pairs~\cite{yin2020novel}, improve speaking style in text-to-speech~\cite{li2022enhancing}, and model users and videos in social media~\cite{wei2019mmgcn}. Inspired by this progress, TMac~\cite{liu2023tmac} applies graph learning to acoustic events by representing audio and video segments as nodes and their temporal links as edges. However, most graph-based methods still treat intra- and inter-modal relations equally, ignoring their temporal differences. This often results in poor alignment and limited representation power.

To address these limitations, we propose a new framework called Temporally Heterogeneous Graph-based Contrastive Learning (THGCL). Our method explicitly models temporal relations between nodes by assigning different weights to intra-modal and inter-modal interactions based on their temporal consistency. We employ Gaussian processes to capture smoothness within each modality and Hawkes processes to model decaying dependencies across modalities. In addition, we introduce a contrastive learning objective to better reduce cross-modal noise. Through this design, THGCL improves both the robustness and expressiveness of multimodal representations. We evaluate THGCL on the AudioSet. The results show that our method achieves state-of-the-art performance.

\section{Our Approach}
\label{sec:approach}


We propose a new framework called THGCL, shown in Fig.\ref{fig:fig1}. The THGCL constructs a temporal heterogeneous graph that models both temporal and semantic dependencies across modalities for acoustic event classification. 

\subsection{Feature encoding module}
\label{ssec:fem}

We begin by extracting audio and visual features to build the foundation of our framework. 
For audio, we first extract clips to 960 ms segments and convert segments to the $96\times 64$ log-mel spectrogram.
This spectrogram is passed to the VGGish~\cite{VGGish}, producing 128-dimensional audio features. For video, each clip is partitioned into non-overlapping 250 ms chunks. Each chunk is processed by the pre-trained S3D network \cite{xie2018rethinking,han2020self}, yielding 1024-dimensional video features.

To align the dimensions, we apply linear transformations to produce embeddings $E_a$ and $E_v$ with the dimension $d$ of both embeddings. These aligned embeddings serve as the inputs for graph construction subsequently.

\subsection{Construction of a temporal heterogeneous graph}
\label{ssec:THG}
We represent each audiovisual input as a temporal heterogeneous graph (THG)\cite{liu2023tmac,zhang2019heterogeneous,amir2022visually}. Here we provide a comprehensive description of the construction process for one graph $G$. We first divide the acoustic and visual embeddings into $P_a$ and $P_v$ segments, respectively. The node sets of \(G\) consist of acoustic nodes $\mathbb{V}^a=\{v_{1}, ..., v_{P_a}\}$ and visual nodes $\mathbb{V}^v=\{v_{1}, ...,v_{P_v}\}$, while the edge set includes intra-audio, intra-visual, and inter-modal edges with adjacency matrices $A_a$, $A_v$, and $A_{inter}$. The intra- and inter-modal interactions that define the edges in the graph are controlled by two parameters, dilation and span across time. For acoustic nodes, the fine granularity of audio features requires precise temporal alignment. For visual nodes, the interactions are influenced by spatial resolution and motion, which allows coarser temporal alignment. For edges across modalities, temporal coherence must be strictly followed, and edges exist only when timestamps match. Once the temporal heterogeneous graph is constructed, the resulting representation becomes consistent with the original audiovisual data. Temporal weights are then applied to refine the edges.

\textbf{For intra-modal relations}, we use a Gaussian process~\cite{fang2021gaussian,li2023heterogeneous} to ensure that nodes closer in time receive higher weights. The weighted adjacency matrices for audio and video are defined as \(\bar{A}_{a}\) and \(\bar{A}_{v}\) with temporal weights for modality \(m\):


\begingroup
\vspace{-6mm}
\small
   \begin{equation}
s^{i,j}_{m} =\exp\left(-\frac{||P_m^{i}-P_m^{j}||^{2}}{2\times(P_m^{\max}-P_m^{\min}+1)^{2}}\right), m\in \{a,v\},
\end{equation}
\vspace{-4mm}
\endgroup


\noindent Here, $P^{\max}$ and $P^{\min}$ represent the maximum and minimum segment index within the neighborhood of the current node. These temporal weights reflect the smoothness of relationships in the embedding space, where nearby segments tend to be more similar in both semantics and features.

\textbf{For inter-modal relations}, the representations of audio and video often differ, and it is necessary to model the decaying effect of past interactions. We apply the Hawkes process \cite{hawkes1971point} to weight these edges, which assigns stronger influence to recent interactions. The inter-modal adjacency is defined as \(\bar{A}_{inter}\) with the temporal weights:


\begingroup
\vspace{-7mm}
\begin{equation}
s^{i,j}_{inter} =\sigma \left ( \frac{\log\xi}{\log(1-\xi)} + \frac{P^{\max'}_{a}-P^{i}_{v}+1}{P^{\max'}_{a}-P^{\min'}_{a}+1} \right ) /\tau,
\end{equation}
\vspace{-4mm}
\endgroup


\noindent Here, $\xi\sim U(0,1)$ introduces randomness, $\tau$ controls smoothness, and $\sigma(\cdot)$ is the sigmoid function. The weights reflect the stronger effect of nodes that are closer in time, while distant events gradually lose their influence. By combining these strategies, we obtain the weighted adjacency matrices ${\bar{A}_a, \bar{A}_v, \bar{A}_{inter}}$, which preserve temporal coherence across modalities. These matrices are then used as inputs to the graph neural network. In practice, audiovisual inputs are processed in batches, which allows parallel training and improves scalability.

\subsection{Model training scheme}
\label{ssec:THGN}



\begin{figure}[t!]
\centering
\includegraphics[width=0.68\columnwidth]{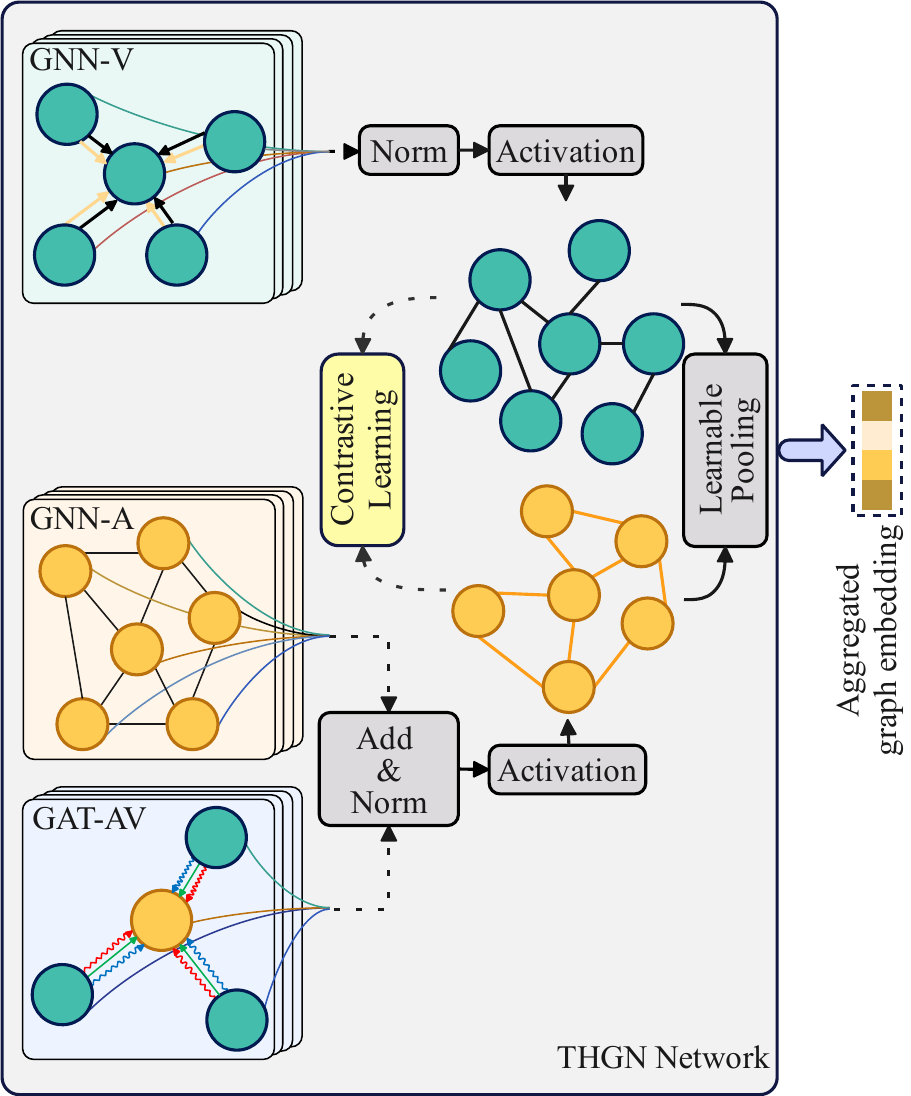}
\vspace{-6pt}
\caption{Architecture of THGN network.}
\label{fig:fig2}
\vspace{-14pt}
\end{figure}


To learn node representations and edge weights across modalities, we propose a Temporal Heterogeneous Graph Network (THGN). The goal is to reduce the influence of visual features that may not support acoustic classification, and second, to filter out irrelevant acoustic details through graph aggregation.

Given embedding set $E=\{E_a,E_v\}$ and weighted adjacency matrices set $\bar{A}=\{\bar{A}_{a},\bar{A}_{v},\bar{A}_{inter}\}$, the GNN encoder can be written as $X^{out}=\Gamma(E,\bar{A},\Psi)$, where $\Psi$ denotes the learnable parameters. The $l$-th layer is defined as:

\begingroup
\vspace{-3mm}
\begin{equation}
X^{l} = \rho(\bar{A}X^{l-1}\Psi^{l-1}), l=
1,...,\mathbb{L},
\end{equation}
\vspace{-6mm}
\endgroup

\noindent with $X^0=E$ as the input, $X^{out}=X^{\mathbb{L}}$ as the final output, and $\rho(\cdot)$ as a non-linear activation such as ReLU.

THGN aggregates intra- and inter-modal information while respecting temporal weights. As shown in Fig.\ref{fig:fig2}, it uses four layers to process audio and visual inputs, then integrates video information into audio nodes. This ensures the model remains focused on acoustic events while benefiting from visual cues. The network includes four temporal graph layers, a contrastive module, and a learnable pooling layer. Audio nodes ($X_a^l$) are updated by GNN-A, video nodes ($X_v^l$) by GNN-V, and cross-modal transfer is achieved through GAT-AV.

To further improve representation and reduce the noisy effects across modalities, we add a contrastive module. Unlike standard multi-modal contrastive learning, our approach works at the clip level with heterogeneous graphs. We design a self-discrimination task that enforces higher similarity across modalities for the same target than for different targets within the same modality. Inspired by SimCLR \cite{chen2020simple}, we define the loss from video to audio accordingly.

\begingroup
\vspace{-5mm}
\small
\begin{equation}
L_{v\to a}=- \frac{1}{\mathbb{B}}\sum_{i\in\mathbb{B}} \log{\frac{\exp \left ( \cos \left ( X_{a}^{(i)}, X_{v}^{(i)} \right ) / \mathfrak{t} \right )}{\sum_{j\in \mathbb{B},j\ne i}\exp \left ( \cos \left ( X_{a}^{(i)}, X_{v}^{(j)} \right ) / \mathfrak{t} \right )} },
\end{equation}
\vspace{-3mm}
\endgroup

\noindent where $\mathbb{B}$ denotes the training batch, $X_{a}^{(i)}$ and $X_{v}^{(i)}$ are the graph embeddings of the $i$-th sample, $\cos(\cdot)$ is the cosine similarity, and $\mathfrak{t}$ is the temperature parameter. A symmetric loss $L_{a\to v}$ is defined from audio to video, and the final contrastive loss $L_{CL}$ is the sum of $L_{a\to v}$  and $L_{v\to a}$.


Then we next connect it to the classification task. Since THGCL is designed for acoustic event classification, we apply a learnable pooling function \cite{liu2023tmac} to aggregate the outputs $X^{out}_{a}$ and $X^{out}_{v}$ into a graph-level embedding $X^{G}$. This embedding is passed into a classification head, which is trained with focal loss \cite{lin2017focal} to address class imbalance. The overall training objective of THGCL is then formulated as a weighted combination of classification loss and contrastive loss:

\begingroup
\vspace{-4mm}
\begin{equation}
L=\omega_{FL} L_{FL} + \omega_{CL} L_{CL},
\end{equation}
\vspace{-6mm}
\endgroup

\noindent where $\omega_{FL}=1.0$ and $\omega_{CL}=0.1$ in our experiments. This joint objective balances accurate event classification with robust cross-modal representation learning.

\section{Experiments}
\label{sec:exp}

\subsection{Experimental settings}
\label{ssec:exps}

\subsubsection{Dataset}
\label{sssec:data}

Our experiments are conducted on the AudioSet \cite{audioset}. Each audio clip has a fixed duration of 10 seconds. We construct a high-confidence subset by selecting 33 sound event categories whose rater confidence scores fall within the range $\left [ 0.7,1.0 \right ]$, resulting in $82,410$ audio-visual training samples. We evaluate our method on the original evaluation set, which contains $85,487$ test clips \cite{liu2023tmac} for fair comparison.

\subsubsection{Implementation and metrics}
\label{sssec:imple}

In the feature encoding module, we set the transformation dimension $d$ as 128. For temporal heterogeneous graph construction, the temporal spans between audio, video, and inter-modal nodes are set to 6, 4, and 3, respectively, while the intra-modality temporal dilations are set to 3 and 4. During training, we adopt the Adam optimizer with an initial learning rate of 0.005. The maximum number of iterations is capped at 5000, and early stopping is enabled. The THGN network has a hidden channel size of 512. Performance is evaluated using mean average precision (mAP) and the area under the ROC curve (AUC). Our code is released at this repository \footnote{https://github.com/visionchan/THGCL.git}.



\subsection{Overall comparison}
\label{ssec:comp}

\begin{table}[t!]
\centering
\caption{The comparable study of Acoustic event detection performance in Audioset. ``$\star$" means graph-based methods.}
\label{tab:model_comparison}
\resizebox{!}{0.45\columnwidth}{
\begin{tabular}{l|ccc}
\toprule
Model & mAP (\%) & AUC & \#Params (M) \\
\midrule
\cellcolor[HTML]{dcdadd}THGCL (Ours) & \cellcolor[HTML]{dcdadd}\textbf{57.4} & \cellcolor[HTML]{dcdadd}\textbf{0.948} & \cellcolor[HTML]{dcdadd} 4.8 \\
TMac $\star$ \cite{liu2023tmac} & \underline{55.1} & 0.937 & 4.3 \\
VAED $\star$ & 51.6 & 0.919 & 2.1 \\
PaSST-S & 49.0 & 0.900 & 87.0 \\
ASiT \cite{asit} & 48.5 & -- & 85.0 \\
Audio-MAE (local) & 48.2 & -- & 86.0 \\
ATST-clip \cite{ATST} & 47.8 & -- & 86.0 \\
MaskSpec & 47.3 & -- & 86.0 \\
LHGNN $\star$ \cite{LHGNN} & 46.6 & -- & 31.0 \\
Conformer-based \cite{conformer-based} & 44.4 & -- & 88.0 \\
HGCN $\star$ & 44.2 & 0.885 & 42.4 \\
Wave-Logmel & 44.1 & -- & 81.0 \\
AST & 44.0 & -- & 88.0 \\
SSL graph $\star$ & 43.9 & -- & 0.2 \\
Wav2vec2-audio & 42.6 & 0.880 & 94.9 \\
VATT & 39.7 & -- & 87.0 \\
ResNet-1D-both & 38.0 & 0.891 & 81.2 \\
ConvNeXt-femto \cite{ConvNeXt-femto} & 37.9 & -- & 5.0 \\
R(2+1)D-video & 36.0 & 0.810 & 33.4 \\
ResNet-1D-audio & 35.9 & 0.900 & 40.4 \\
\bottomrule
\end{tabular}}
\vspace{-4mm}
\end{table}

Table \ref{tab:model_comparison} shows that the proposed THGCL achieves the best mAP of 57.4\% and AUC of 0.948 with only 4.8M parameters, showing both accuracy and efficiency. The poor results of non-graph-based methods at the bottom of the table \ref{tab:model_comparison} highlight the importance of leveraging graph structures to model complex temporal relationships for acoustic event classification. Heterogeneous graph-based models such as TMac and VAED capture structural relations but ignore temporal alignment and decays, limiting their gains. Waveform models such as Wave-Logmel, and large ResNet-1D variants, struggle with noise sensitivity and inefficiency. Consequently, they exhibit poor performance. In contrast, THGCL unifies Gaussian and Hawkes weighting with contrastive learning, capturing intra- and inter-modal temporal dependencies while suppressing noise, which leads to state-of-the-art results beyond all baselines.




\subsection{Ablation studies of relation modeling strategies}
\label{ssec:abla}

\begin{table}[t!]
\centering
\caption{Performance comparison of different temporal types}
\label{tab:temporal_comparison}
\begin{tabular}{c|l|cc}
\toprule
ID & Temporal type & mAP (\%) & AUC \\
\midrule
1 & \cellcolor[HTML]{dcdadd} w/ Gau. \& Haw. & \cellcolor[HTML]{dcdadd}\textbf{57.4} & \cellcolor[HTML]{dcdadd}\textbf{0.948} \\
2 & both Haw. & 55.0 & 0.942 \\
3 & both Gau. & 53.5 & 0.893 \\
\bottomrule
\end{tabular}
\vspace{-2mm}
\end{table}

\begin{figure}[t!]
\centering
\includegraphics[width=1\columnwidth]{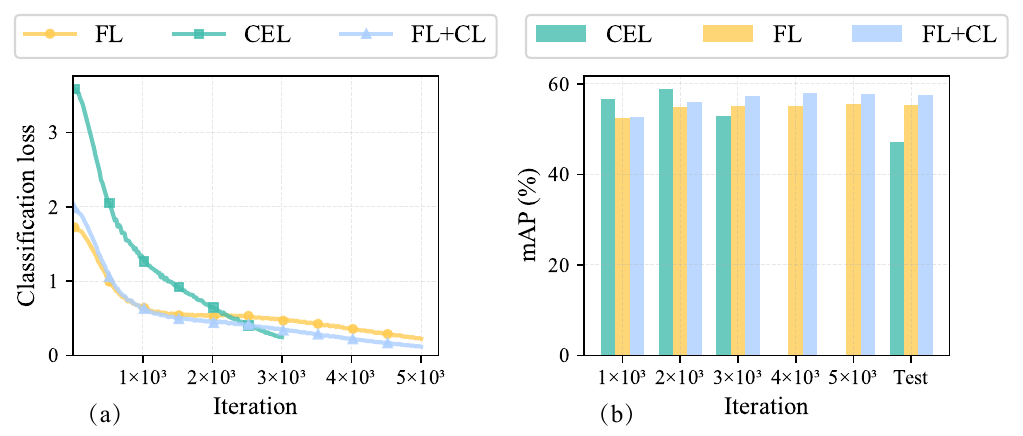}
\vspace{-16pt}
\caption{Compare effect of different loss functions for acoustic event classification. (a) Training loss versus iteration. (b) Performance changes over different stages.}
\label{fig:fig3}
\vspace{-10pt}
\end{figure}

To assess the effectiveness of the temporal graph construction on the model's discriminative capacity, we first compare alternative temporal relation modeling strategies in Table \ref{tab:temporal_comparison}. Specifically, ID-1 denotes our configuration mentioned in section \ref{ssec:THG}; ID-2 replicates the TMac setting; ID-3 applies the Gaussian process (Gau.) to all temporal edges. The results show that ID-1 achieves the best performance. Using Gau. within the intra-modality is effective because acoustic events exhibit short‑term stationarity and adjacent nodes share similar representations, employing Hawkes process (Haw.) across modalities better characterizes cross‑modal excitation and alignment, thereby preserving temporal consistency and enhancing overall discriminative performance.

\subsection{Ablation studies of different loss functions}
We also explore the contribution of self‑supervised contrastive learning by exploring different loss functions.
Fig. \ref{fig:fig3}(a) shows that cross entropy loss (CEL) converges the slowest and remains at a higher loss; focal loss (FL) descends faster early on but soon plateaus; our loss $L$ (FL+CL) achieves both rapid initial decrease and continued refinement, reaching the lowest and smoothest final loss. Fig. \ref{fig:fig3}(b) indicates that FL surpasses CEL in early mAP, while FL+CL leads throughout and sustains a stable mid‑to‑late advantage. This demonstrates that incorporating a contrastive loss enables the model to learn stronger cross‑modal representations and reduces the impact of noise contamination, enhancing generalization and mitigating overfitting. Overall, the joint loss surpasses classification losses in convergence speed, final performance, and stability.

\section{CONCLUSION}
\label{sec:typestyle}

We propose a Temporal Heterogeneous Graph Contrastive Learning (THGCL) method to cross-modal temporal misalignment and background noise interference in acoustic event classification. THGCL builds a temporal heterogeneous graph, where intra-modal edges use Gaussian weights for smooth coherence and inter-modal edges use Hawkes weights for synchronization. A Temporal Heterogeneous Graph Network (THGN) then combines contrastive learning with embedding aggregation to enhance cross-modal consistency, enlarge class margins, and suppress noisy segments. Experiments on AudioSet show that THGCL outperforms strong baselines and that its modules are complementary.

\clearpage
\footnotesize
\bibliographystyle{IEEEbib}
\bibliography{strings,refs}

\end{document}